\documentclass{rspublic}
\usepackage{epsfig}
\begin{document}

\newcommand{\apj}{{\it Astrophys. J.}}
\newcommand{\aj}{{\it Astron. J.}}
\newcommand{\mnras}{{\it Mon. Not. R. Astron. Soc.}}
\newcommand{\aanda}{{\it Astron. Astrophys.}}
\newcommand{\pasp}{{\it Publ. Astron. Soc. Pac.}}
\newcommand{\ptra}{{\it Phil. Trans. R. Soc. A}}

\title[Chemical evolution of star clusters]{Chemical evolution of star
       clusters}
\author[J. Th. van Loon]{Jacco Th. van Loon}
\affiliation{Lennard--Jones Laboratories, Keele University, Staffordshire ST5 5BG, UK}

\label{firstpage}

\maketitle 

\begin{abstract}{{\bf stars: abundances; stars: evolution; Galaxy:
evolution; Galaxy: formation; globular clusters: general; open
clusters and associations: general}} I discuss the chemical evolution
of star clusters, with emphasis on old globular clusters, in relation
to their formation histories. Globular clusters clearly formed in a
complex fashion, under markedly different conditions from any younger
clusters presently known. Those special conditions must be linked to
the early formation epoch of the Galaxy and must not have occurred
since.  While a link to the formation of globular clusters in dwarf
galaxies has been suggested, present-day dwarf galaxies are not
representative of the gravitational potential wells within which the
globular clusters formed. Instead, a formation deep within the
proto-Galaxy or within dark-matter minihaloes might be favoured. Not
all globular clusters may have formed and evolved similarly. In
particular, we may need to distinguish Galactic halo from Galactic
bulge clusters.
\end{abstract}

\section{Prologue}

Clusters of stars are ideal witnesses of past star formation and
chemical evolution. This is because the stars in a cluster outline a
sequence in photometric diagrams which can be compared with
theoretical models for stellar evolution to infer an age and
metallicity for the cluster. This way, young ($<100$ Myr) clusters can
be used to investigate the mechanisms of triggering and propagation of
star formation in detail and the dependence of star formation on local
environmental conditions. Intermediate-age ($\approx0.1$--10 Gyr)
clusters trace much of the history of galaxies, while old ($>10$ Gyr)
globular clusters (GCs) probe the formation epoch of galaxies.  The
`chemical' (really: `elemental') composition of a cluster is a
powerful legacy of the entire history of star formation, stellar
evolution, dynamical and interstellar medium (ISM) processes
pre-dating the formation of the cluster. The cluster age tags this
imprint to a specific time in the history of a galactic system.

Complications arise from the fact that the position and kinematics of
a cluster may no longer be directly connected to its place of origin
and birth kinematics. In the Milky Way, dynamical heating leads to
outward migration away from the disc midplane and the Galactic
Centre. Clusters dissolve over time, and those that survive for
many billions of years generally have masses around $10^4-10^6$
M$_\odot$. Their formation requires conditions which are uncommon in
the present, nearby Universe. In addition, attempts to quantify the
star-formation rate are hampered by the uncertain relationship between
the conditions under which stars were formed and the number and size
of clusters which were produced. Nevertheless, their chemical
composition can constrain all of these open issues.

I concentrate on new lines of evidence, emerging in a fast-paced field
of research, complementing and complemented by the excellent reviews
of Gratton {\it et al.}\ (2004) and Friel (1995), Freeman \&
Bland--Hawthorn (2002) and Brodie \& Strader (2006).

\section{The agents of chemical evolution}

We must first define what we mean by metallicity. In astrophysics,
metals are elements heavier than hydrogen (H), helium (He), lithium
(Li) and beryllium (Be). The metallicity is the fraction, in mass, in
all these elements combined. This amounts to at most $Z\approx0.05$,
is Z$_\odot\approx0.02$ for the Sun, and can be as little as
$Z\approx0.0001$ in the most metal-deficient Galactic GCs (and less
still in extremely metal-deficient halo field stars).  In most
main-sequence (MS) stars, He takes up $Y\approx0.25$ of the total,
with the remaining $X\approx0.7$ in H (where $X+Y+Z=1$). Many metals
are very rare and make a negligible contribution to the total. It
would be cumbersome and inaccurate to have to measure the abundances
of individual metals and add them all up to arrive at a value for
$Z$. It is therefore sensible to pick one metal and use that as a
proxy for metallicity. Often, iron (Fe) is used, relative to that of H
and benchmarked against the Sun, denoted as [Fe/H]. Fe is common, and
its production in supernovae (SNe) results in a steady increase over
time as subsequent generations of stars incorporate the accumulated
enrichment by the previous generations. This scenario may, in fact, be
traced more promptly by oxygen (O), which is even more common and
relatively easily measured in the ISM of other galaxies. Yet,
different histories will lead to different relative enhancements (or
even depletion) of the various elements, including O and Fe, so there
really exists no true single chemical yardstick.

\subsection{The origin of the elements}

Elements heavier than H are synthesized both in stars and in
explosions.  Differences in yields are generally associated with the
temperature or neutron density at which nuclear synthesis takes
place. But to speak of a yield in terms of chemical enrichment of the
Universe, the products must leave the gravitational boundary of their
source. In stars, this requires a mechanism of transport through the
mantle into the surface layers, followed by mass loss.  In explosive
events, it requires that the products join the ejecta instead of being
encapsulated within a compact remnant (neutron star, NS, or black
hole).  The yields from stars generally depend on other parameters in
addition to mass and composition, for instance rotation and magnetic
fields, which both affect the production rates and mixing efficiencies
and possibly also the mass-loss rate.  The physics of mixing and mass
loss are not fully understood, and are only incorporated in stellar
models in an ad hoc, parameterized way. And some crucial reaction
rates are uncertain by more than an order of magnitude.

The lifetime of a star is roughly also the delay timescale between its
formation and its contribution to chemical enrichment of the
ISM. Chemical enrichment of subsequent generations of stars is further
delayed between injection into the ISM and participation in star
formation. Table 1 gives a succinct overview of the main protagonists,
the (minimum) timescales on which they operate and the key
contributions they make to chemical enrichment.

\begin{table}
\caption{Factories of the elements}
\begin{tabular}{lccl}
\hline
Source & Birth mass (M$_\odot$) & Timescale & Key elements \\
\hline
Massive stars & $>20$    & 2--10 Myr   & He, N (rotating, or Wolf--Rayet-type) \\
SN II         & 8--20    & 10--40 Myr  & He, CNO, Fe, Si, Ca, r-process \\
AGB (HBB)     & 4--8     & 40--200 Myr & N, He, Na, s-process \\
SN Ia         & 1--8 ?   & $\sim$ Gyr  & Fe, Cu \\
AGB (C)       & 1.5--4   & 0.2--2 Gyr  & C, F, s-process \\
RGB           & 0.8--1.5 & 2--12 Gyr   & $^{13}$C, $^{14}$N \\
\hline
\end{tabular}
{\scriptsize AGB: asymptotic giant branch, HBB: hot-bottom burning, RGB: red-giant branch.}
\end{table}

\subsubsection{Fe-peak elements}

These elements, with atomic masses near that of Fe, are ejected in
SNe. In particular the presence of nickel (Ni) is a tell-tale
signature of such events, with $^{56}$Ni the product of the final
nuclear burning stage (that of silicon, Si, by $\alpha$ capture) in
massive stars ($>8$ M$_\odot$). The decay of $^{56}$Ni to the cobalt
isotope $^{56}$Co and subsequently to $^{56}$Fe creates much of the Fe
released in SNe. In Type II SNe, much of the Ni and Fe is locked up in
the collapsing core, but very massive ($>130$ M$_\odot$) metal-poor
stars explode entirely as a result of pair creation, leaving no
remnant. Fe is also produced in large quantities in explosive
nucleosynthesis in Type Ia SNe, which are devoid of H as they are
detonations or deflagrations of white dwarfs (WDs). All this Fe ends
up in the ISM. Because WDs are the products of intermediate-mass stars
(0.5--8 M$_\odot$), enrichment by SNe Ia is slower ($10^8-10^9$ yr;
Raskin {\it et al.}\ 2009) than that by SNe from massive stars. A Type
Ia SN likely results from the WD having grown to the Chandrasekhar
mass limit (1.4 M$_\odot$) by accreting matter from a companion star,
or from the merger of two WDs. This additional delay is very
uncertain: massive WD progenitors leave WDs with masses closer to the
Chandrasekhar limit, but the mutual separation of the mass donor and
receiving WD (or of the two WDs), and hence the efficiency of mass
transfer (or orbital degradation), may also depend on the birth mass.

The net result is that the Galactic disc grew to [Fe/H] $\approx-1$
dex through SNe from massive stars, after which Type Ia SNe kicked in
on a typical delay timescale of $\sim$ Gyr. Subsequently, the
continued increase in Fe content diluted earlier enrichment from
massive stars in other elements (`Q'), decreasing their [Q/Fe] ratios
towards zero (solar). Copper (Cu) and manganese (Mn) are produced more
lavishly in Type Ia SNe. Hence, they are underabundant with respect to
scaled-solar abundances by 0.3--0.7 dex at [Fe/H] $<-1$ dex, and only
reach solar values after prolonged enrichment by Type Ia SNe. Other
Fe-peak elements are titanium (Ti), which is easy to measure in
stellar spectra, and vanadium (V), which forms easily recognizable
molecules in the atmospheres of the coolest M-type giant stars, and
chromium (Cr).

\subsubsection{The r-process}

Heavy elements are produced by neutron (n) capture, starting with Fe
(the `seed' nucleus). This proceeds rapidly in the explosive deaths of
massive stars, and anything as massive as uranium (U) can result. An
excellent tracer of this process is europium (Eu). Indeed, in the
Galaxy, [Eu/Fe] $\approx0.4$ dex up to [Fe/H] $\approx-1$ dex, after
which it decreases towards solar values, essentially showing the
opposite behaviou is produced through $\alpha$ capture. C (and
s-process elements) is transported to the surface as the convective
envelope penetrates the production site as a result of the switch
between H- and He-shell burning (thermal pulsing, TP). In
subsolar-metallicity stars, this turns the star into a C star as the
surface C/O ratio exceeds unity. In massive AGB stars (4--8
M$_\odot$), the bottom of the convection zone is so hot that an
incomplete CNO cycle operates (hot-bottom burning: HBB): C is
converted into N, but not subsequently into O. These stars produce N
and deplete C. Isotopic changes occur too: for instance, $^{18}$O is
depleted and the $^{12}$C/$^{13}$C ratio is reduced. Na, aluminium
(Al) and Mg can be synthesized as well {\it through} p-capture,
especially at ever lower metallicities: at very low metallicities,
$Z\sim10^{-4}$, in the most massive AGB stars, O is more fully
depleted than C, and these stars become C stars even though they do
not enrich the Universe with C but instead with large amounts of N
(Ventura \& D'Antona 2009).

\subsubsection{Helium}

He is produced in all stars $>0.5$ M$_\odot$, but it does not
necessarily end up in the ISM. One would expect massive AGB stars
undergoing HBB to yield He.  Super-AGB stars, which behave like AGB
stars but proceed to burn C in the core, contribute He as well as
enhancing C+N+O (Pumo {\it et al.}\ 2008).  Their fate as either ONe
WD or electron-capture SN is unclear. Prolific producers are
Wolf--Rayet (WR) stars. These have shed their H envelope as a result
of strong stellar winds in their previous phase as an O- or B-type
supergiant or possibly after an intermediate phase as a red supergiant
(RSG).  Their surface layers consist predominantly of He, with either
abundant N (WN; $20<M<40$ M$_\odot$) or C and O (WC; $>40$
M$_\odot$). WR stars also have strong winds and thus enrich the ISM
with He, as well as N, or C and O. Note, however, that if one adds up
the lost H, one arrives at more similar enrichments to those
contributed by SNe of Type II ($8<M<20$ M$_\odot$).

\subsubsection{Fluorine}

Intermediate in mass between O and Ne, fluorine (F) is produced in the
OF-cycle variant of the CNO cycle, most likely in intermediate-mass
TP--AGB stars, as it is destroyed in more massive stars (Lebzelter
{\it et al.}\ 2008; Abia {\it et al.}\ 2009).

\subsection{Internal chemical evolution}

Care must be taken not to mistake internal chemical processes,
happening within the cluster since its formation, for primordial
enrichment. This is a prime concern in old clusters, where the evolved
giant stars are so much brighter and hence more readily accessible to
spectroscopic investigation than their MS siblings. Such interfering
processes can be internal to the star itself, or internal to the
cluster but external to the star in question.

\subsubsection{Surface abundances affected by stellar evolution}

In cool giants, `mixing' is due to convection, the mode of energy
transport from the nuclear production site to the
photosphere. Normally, a radiative layer exists between the nuclear
burning site and the convective mantle, acting as a buffer. But
occasionally, restructuring of the mantle as a consequence of
switching nuclear furnace allows the convection zone to penetrate,
leading to a `dredge-up' episode. Of particular intensity, the third
dredge up occurs repeatedly (briefly, with intervals of typically
$10^4$ yr) during the TP--AGB phase. This brings C and s-process
elements to the surface, creating a C star (or an S-type star which
has C/O$\approx1$ and is thus rare) or a N-enriched star if HBB
operates. It is accompanied by short-lived elements such as technetium
(Tc), so if Tc is seen then the surface enrichment is
recent. Especially in metal-poor intermediate-mass stars, the C/O
ratio on the surface is easily reverted from O to C dominated. Because
the molecular chemistry of the atmospheres of cool stars ($T<4000$ K)
is largely determined by whatever element is left after formation of
carbon monoxide (CO), C stars look very different from O-rich red
giants, displaying strong C$_2$ and TiO bands, respectively. In cool
stars with C/O$<1$, carbonaceous molecules are still present, most
notably CH (observable as, e.g., the `G band' at 4300 \AA) and CN
(e.g., around 3800 \AA).

In low-mass stars ($<1.5$ M$_\odot$), during their ascent of the
red-giant branch (RGB), CNO-cycle burning in the H shell can drive CNO
isotopes to their equilibrium values, which are very different from
their primordial levels: $^{12}$C/$^{13}$C $\rightarrow3.5$ (during
AGB H-shell burning this ratio is typically around 15--20, with a
solar value of $\approx90$) and $^{14}$N and $^{17}$O become abundant
isotopes. The first dredge up on the RGB is not sufficiently efficient
and perhaps absent in metal-poor stars. A noncanonical mixing
mechanism must be operating above the RGB `bump' (around $M_V\approx0$
mag, caused by a discontinuity in the molecular weight). Rotation, or
the thermohaline effect (a molecular-weight inversion causing
buoyancy; Eggleton {\it et al.}\ 2006; Charbonnel \& Zahn 2007), have
been suggested as its origin. Whatever its cause, it must be more
effective at lower metallicity, as $^{12}$C/$^{13}$C ratios close to
the equilibrium value are observed in metal-poor, luminous RGB stars,
[Fe/H] $<-0.8$ dex, but not in their more metal-rich counterparts
(Charbonnel \& Do Nascimento 1998).

Li is normally depleted during the first dredge-up episode: Pasquini
{\it et al.}\ (2005; in NGC\,6752) and Bonifacio {\it et al.}\ (2007;
in 47\,Tuc) provide evidence for a Li--Na anticorrelation, consistent
with the admixture of gas depleted in Li and rich in Na produced by
p-capture reactions. However, the anticorrelation appears in GCs but
not in the field, and may thus result from the GC formation
mechanism. The Galactic GC NGC\,6397, on the other hand, exhibits a
uniform Li content across its unevolved population. Rare super-Li-rich
stars have been found, too, and some unconventional process is
required to explain this happening some way up the RGB (Abia {\it et
al.}\ 1991). Li enhancement in Galactic massive AGB stars does not
appear to be accompanied by s-process enhancement
(Garc\'{\i}a--Hern\'andez {\it et al.}\ 2007), while in the Magellanic
Clouds it is (Smith {\it et al.}\ 1995): a metallicity effect.

In stars with radiative mantles, other effects can cause modification
of the surface chemical abundances. In horizontal-branch (HB) stars
(post-RGB, core-He burning giants) gravitational settling and
selective radiative levitation can lead to differentiation of metals,
making these stars highly unsuitable for studying primordial
enrichment or even later enrichment by external processes. This is a
pity as, for instance, He lines are visible in the spectrum of stars
in this temperature regime ($T>11\,500$ K). More stars of spectral
type A display peculiarities (`Ap' stars), some believed to be due to
the acting of magnetic fields (metallic, `Am' stars).

Fast rotation may also induce mixing of the star where it would
otherwise not be expected. This might produce hot stars with a surface
enriched in N, as well as C and O. This contributes to the uncertainty
about the main producers of N in the early Universe: rotating massive
stars (Decressin {\it et al.}\ 2007) or perhaps massive AGB stars
(Ventura \& D'Antona 2009)?

Finally, surface chemical abundances may undergo dramatic changes as
the star sheds its H-rich mantle, directly exposing the layers
underneath in which nuclear burning had taken place previously. This
gives rise, for instance, to post-AGB objects enriched in s-process
elements. Or massive WR stars, which are easily recognized by their
very broad emission lines (no absorption lines at all). The presence
of WR stars in a cluster is related to the cluster's age, but may be
affected by additional properties of the stars such as their
metallicity at birth and rotation rate. Similarly, C stars in a
cluster help constrain its age.

\subsubsection{Accretion}

In dense stellar systems, external pollution by other stars may be
relatively important: accretion from a pool of gas, possibly
replenished by the winds from other stars, or mass transfer from
binary companions, or {\it through} stellar mergers.

In a standard Bondi--Hoyle--Lyttleton scenario, the accretion rate from
diffuse gas is remarkably well approximated by
\begin{equation}
\dot{M}=\frac{4\pi G^2M^2\rho}{(c_{\rm s}+v)^3},
\end{equation}
where $M$ is the mass of the accreting star, $\rho$ the density of the
medium it passes through, and $c_{\rm s}$ and $v$ are the sound and
bulk speed, respectively. In 10 Gyr, a star of $M=1$ M$_\odot$ moving
through ISM of $\rho=2\times10^{-24}$ g cm$^{-3}$ ($\approx1$ particle
cm$^{-3}$) and $c_{\rm s}=10$ km s$^{-1}$ at $v=10$ km s$^{-1}$ will
accrete just $\Delta M\approx1\times10^{-5}$ M$_\odot$. Denser gas has
not been found in GCs (Freire {\it et al.}\ 2001; van Loon {\it et
al.}\ 2009). Kinematically colder populations, with $v\sim1$ km
s$^{-1}$ only accrete more efficiently from thermodynamically cold gas
($c_{\rm s}\ll 10$ km s$^{-1}$). In any case, one would expect all
stars to participate in this process and become polluted. Thus,
accretion from diffuse gas to explain abundance anomalies in GCs seems
a rather contrived scenario.

Binary mass transfer is observed in a number of types of systems. If a
MS star accretes mass from the mantle or wind from a C-star red-giant
companion, the chemical imprint upon the radiative layers of the MS
star is obvious: the MS star turns into a C star as well, at least in
appearance, along with the usual s-process signatures such as Ba
enhancement. This naturally explains the observations of C stars in
GCs which are too faint for TPs and the third dredge up to operate,
although it is more challenging to explain such C and Ba stars found
at the tip of the RGB, where the convective mantle would mix and
dilute the accreted C with oxygeneous material (van Loon {\it et al.}\
2007). To prove that these kinds of enrichments are neither due to
binary mass transfer nor to internal nucleosynthesis and mixing, a
consistent pattern must be found along the RGB and preferably also on
the MS.

Stellar mergers may result from common-envelope systems in which a
binary star becomes enveloped by the red-giant mantle of one of the
components as it ascends the RGB. If the cores merge, a more massive
star may remain, at a position on the MS higher than the MS turnoff
associated with the evolution of single stars in a coeval
population. These `blue stragglers' could be mistaken for a younger
population, but the absence of corresponding stars at more evolved
stages of evolution rules out this possibility.

\subsection{Material, chemical, radiative and mechanical feedback}

The total amount of mass lost by a star is reasonably well known (see
also Kalirai \& Richer 2010). But the exact timing of the mass loss
determines the ultimate yields (the initial mass function, IMF,
determines the combined yield), while the wind speed and momentum
determine the degree and timescale of mixing with the surrounding ISM
(this also depends on the gravitational field and ISM density).

AGB stars lose mass through slow winds, $\sim10$--20 km s$^{-1}$
(lower speeds at lower metallicity) and metallicity-independent
mass-loss rates of $\sim10^{-7}$ to $10^{-4}$ M$_\odot$ yr$^{-1}$ (van
Loon 2006). These winds are easily retained within massive GCs, but
may escape from clusters lacking sufficient mass (be it baryonic or
dark). RSG winds are relatively slow too ($<40$ km s$^{-1}$). Stars of
$>30$--40 M$_\odot$ do not reach the RSG phase. These O- and B-type
supergiants, and WR stars, or similar kinds of progenitors or
descendants of RSGs, have much faster winds ($\sim10^3$ km s$^{-1}$).

SNe have extremely fast ($\sim10^4$ km s$^{-1}$) and energetic
outflows, which sweep away any gas present in their vicinity. At some
distance it will run out of steam and stall, but the combined effect
of several SNe creates a large cavity in the ISM. Various types of SNe
are still poorly understood, for instance Types Ib/c (probably
resulting from WR stars) and II--L. The fate of super-AGB stars is
unknown. If they blow up they may efficiently spread their enriched
material over large distances, but otherwise enrichment of the ISM
takes place mainly through their slow winds.

Strong stellar winds and SNe are generally found to compress the
surrounding ISM, and star formation is often seen surrounding the
blown cavities. Similarly, radiative feedback from luminous, hot stars
can induce a wind emanating from a star cluster, and also create an
ionization front which may induce the collapse of molecular cloud
material and thereby trigger star formation. Yet, convincing evidence
of actual triggering of new star formation as a result of the
mechanical or radiative feedback is surprisingly sparse (cf.\ Chu {\it
et al.}\ 2005; Oliveira {\it et al.}\ 2006). The younger generation
does not coincide spatially with the central cluster, and it is
doubtful that the feedback and subsequent star formation (possibly
from chemically enriched gas) ultimately results in a cluster with a
composite population of stars.

\section{Chemical evolution of star cluster systems}

\subsection{The Galactic population of open clusters}

The Galactic system of open clusters (OCs) is fairly well
understood. Their metallicities are around solar or somewhat lower and
they generally compare well to the metal-rich end of the Galactic-disc
field-star population. This is expected, as OCs have ages typically
less than a Gyr and thus formed from material that had been chemically
enriched over many Gyr. The reason for the abundance of such
relatively `young' OCs is not a recent burst in their formation, but
rather the dispersal of older OCs due mainly to evaporation of stars
in combination with continuous dynamical relaxation, and tidal effects
within the Galactic gravitational potential (see also de Grijs 2010;
Larsen 2010).

No age--metallicity relation is present for OCs (cf.\ Pancino {\it et
al.}  2009). This is illustrated in figure 1, where slightly
supersolar-metallicity OCs are seen at all ages, and the clearly
subsolar-metallicity OCs of relatively old age are displaced within
the Milky Way. Indeed, metallicity gradients are seen (Pancino {\it et
al.} 2009; and references therein) with respect to Galactocentric
distance and distance to the Galactic midplane, and similar gradients
are seen in the cluster ages (figure 1). This is not unexpected, as
clusters, like field stars, migrate away from the peak in the mass
density as a result of encounters with gravitationally disturbing
bodies (in the case of OCs, these are likely molecular clouds and/or
spiral arms). The lack of young clusters at large Galactocentric
distances suggests a lack of recent in situ star formation, with
consequentially lower rates of chemical evolution, but this may be
resolved by more systematic searches for young OCs outside the solar
circle.

\begin{figure}
\centerline{\psfig{figure=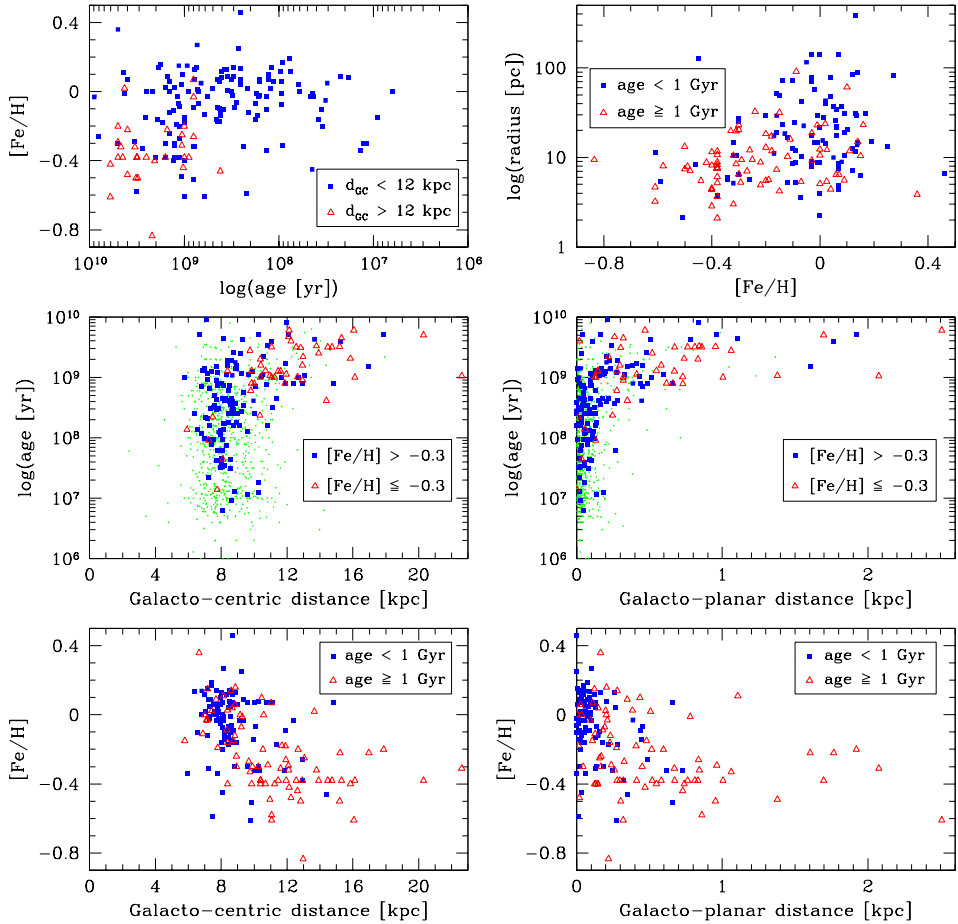,width=126mm}}
\caption{Trends in the global properties of 1787 Galactic open
clusters from the compilation of Dias {\it et al.}\ (2002).}
\end{figure}

The larger OCs have close-to-solar metallicity, with metal-poor
OCs---invariably older---being more compact (figure 1). This may be a
selection effect, as diffuse OCs dissolve over time. Yet massive
($\sim10^4$ M$_\odot$), young ($\sim10^7$ yr) OCs are only found in
the inner regions of the Galaxy, which suggests that their formation
requires stronger gravitational confinement of the molecular cloud
environment. Examples are the well-known Arches and Quintuplet
clusters and Westerlund 1, harbouring very massive stars including WR
stars, and a number of recently-discovered (in infrared surveys) OCs
with large populations of RSGs (Figer {\it et al.}\ 2006; Davies {\it
et al.}\ 2007).  They have near-solar metallicities, as expected from
their age and location in the inner parts of the Galactic disc. (It is
well-known that the Sun is rather metal-rich for its age of 4.6 Gyr,
but also that chemical enrichment in the Galactic disc has proceeded
very slowly, if at all, since the Sun's formation. A reason may be the
ongoing accretion of gas of subsolar metallicity from the halo,
diluting chemical enrichment.)

The massive cluster NGC\,6791 (see also below) is considered an OC,
but being as old and (supersolar) metal-rich as it is, and given its
location in the inner Galactic disc, it is likely associated with the
end of the bulge-formation epoch. Its outward migration has also been
suggested by Boesgaard {\it et al.}\ (2009).

\subsection{The Galactic globular cluster system}

GCs do not contain stars with [Fe/H] $<-2.3$ dex. Stars much more
metal-deficient are found in the halo, so either GCs did not form at
those early, metal-poor times, or they have not survived, either
because they were not massive enough or because they were subjected to
particularly hostile conditions early on. This means that the
present-day, surviving GCs are not the first stellar systems to have
formed, and that their chemical imprint is testimony of the earlier,
`lost' generation (or generations).

The Galactic GC system has long been known to be composed of two main
subsystems, one associated with a halo-like distribution and
kinematics, and the other with a more flattened and kinematically more
ordered distribution: the latter population is referred to as a
`bulge' population. This is illustrated in figure 2, where the
metallicity distribution is clearly bimodal with peaks on either side
of [Fe/H] $\approx-0.9$ dex. It is interesting to note that there is
no correlation whatsoever between mass (by proxy of absolute visual
magnitude, $M_V$) and metallicity. Halo and bulge GCs are similar in
mass.

A change in the HB ratio from positive, if a pronounced hot HB is
present, to negative occurs among the halo GCs (figure 2). It is
uniformly negative among bulge GCs. Between [Fe/H] $\approx-1.3$ and
$-1.8$ dex, GCs are found with both negative and positive HB ratios,
giving rise to the so-called `second-parameter problem'. In fact, the
metallicity at which this occurs is somewhat lower for less massive
GCs (perhaps this is a bias due to the paucity of hot-HB
stars). Indeed, the transition in HB ratio is abrupt for the top two
magnitudes in $M_V$, at [Fe/H] $=-1.3$ dex. The most metal-rich GCs
that show a hot HB have [Fe/H] $=-1.3$ dex even among the less massive
GCs, at $M_V \approx-6$ mag.

Halo GCs extend to lower central luminosity (a proxy for mass) density
than bulge GCs (figure 2). These may be dissolving but given that they
have persisted for so long, a tantalizing alternative is that they
contain a modest amount of dark matter (making up for the lack of {\em
luminous} mass density). There is a trend for higher metallicity at
higher central density, but this is very weak indeed. There does not
appear to be any correlation between metallicity or central density on
the one hand, and ellipticity of the GC on the other, so the amount of
angular momentum inherited from their formation appears to be entirely
random.

The bimodal metallicity distribution is linked to the halo and bulge
division.  The Galactic spatial distributions indeed exhibit a sharp
division, at [Fe/H] $\approx-1.3$ dex when considering Galactocentric
distance or [Fe/H] $\approx-1.4$ dex in terms of Galactoplanar
distance (figure 2), which, curiously, is a few dex lower than the
location of the division in the metallicity distribution. The halo GCs
venture to well over 100 kpc distance, while the bulge GCs stay within
20 kpc from the Galactic Centre. Both systems are flattened, with
Galactoplanar distances being smaller than their Galactocentric
equivalents. Especially the bulge GCs stay within just a few kpc from
the Galactic plane, but halo GCs also often have orbits very close to
the midplane. The less massive GCs, in both systems, are distributed
more widely. This is a clear signature of mass segregation within the
GC systems.

\begin{figure}
\centerline{\psfig{figure=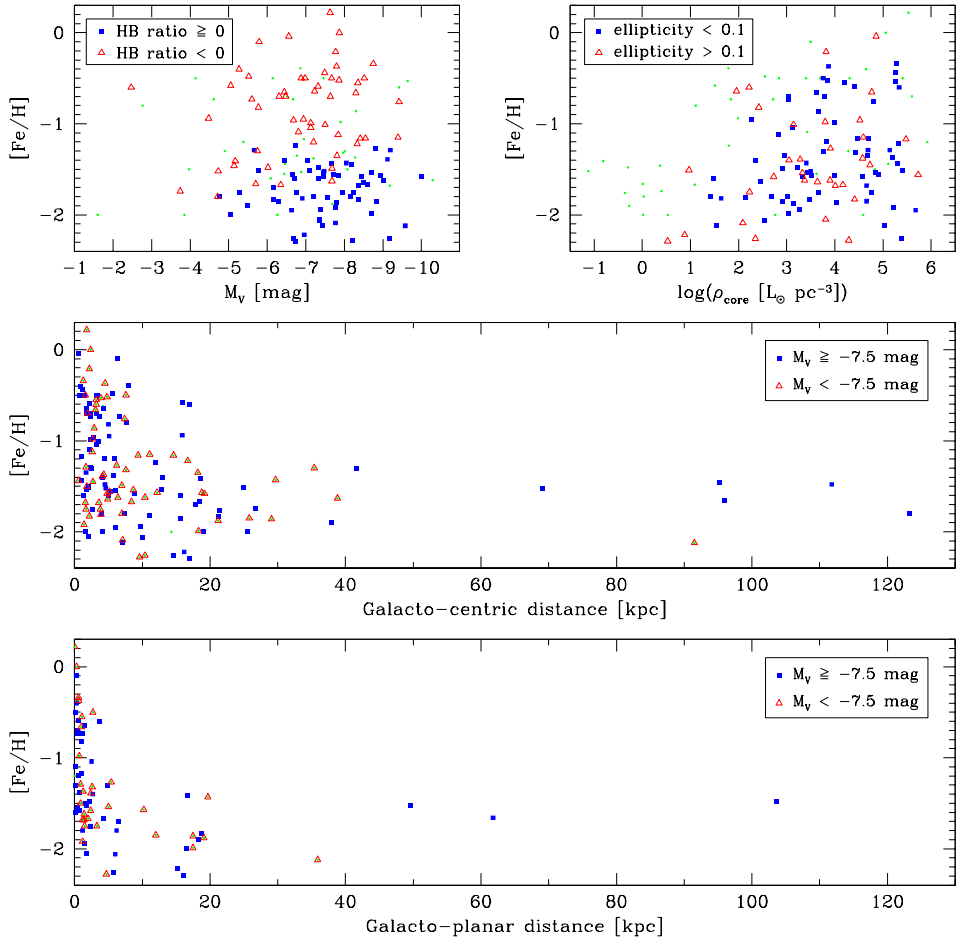,width=126mm}}
\caption{Trends in the global properties of 150 Galactic globular clusters
from the compilation of Harris (1996).}
\end{figure}

Fraix--Burnet {\it et al.}\ (2009), on the basis of a clever
multivariate analysis, deduced an age--metallicity relation for the
halo GCs, from [Fe/H] $\approx-2$ dex 12 Gyr ago to [Fe/H]
$\approx-1.3$ dex 9 Gyr ago. Note, however, that their data are also
consistent with a more rapid increase in metallicity $\approx10$--11
Gyr ago, from [Fe/H] $\approx-1.9$ to $-1.4$ dex. This is also the age
interval in which the bulge GCs seem to have appeared on stage, having
generally higher metallicites---[Fe/H] $\approx-1.4$ to $-0.5$
dex---than halo GCs of similar age.

Fraix--Burnet {\it et al.}\ (2009) further confirmed earlier evidence
that Galactic GCs differ from Local Group dwarf galaxies in having
higher [$\alpha$/Fe] ratios. Halo GCs are indistinguishable from halo
field stars in this respect. Bulge GCs lack the dilution of $\alpha$
elements by further enrichment with Fe from Type Ia SNe that
characterizes field stars in the Galactic disc. This reaffirms the
notion that bulge GCs must have formed, and been chemically enriched,
rapidly.

Sparse data exist for the most metal-rich GCs ([Fe/H] $>-0.5$ dex). In
Liller\,1, NGC\,6553 and NGC\,6528, [Ca/Fe] is as high as for
metal-poor GCs (Gratton {\it et al.}\ 2004) and thus different from
field stars and OCs. There is no dependence of [Ca/Fe] on
Galactocentric distance. NGC\,6528 is also the only metal-rich GC with
measured [Mn/Fe], which is also as low as that in metal-poor GCs. So,
Type II SNe seem to have done the job all the way to [Fe/H] $\approx0$
dex, where in other Galactic components (and possibly in Rup\,106 and
Pal\,12) Type Ia SNe contributed.

Some GCs seem to defy these trends. For instance, Rup\,106 and Pal\,12
are both `young' outer-halo GCs. For their [Fe/H] $\approx-1$ dex,
they have rather low Fe-peak (Ni) and low $\alpha$ (Ca)-element
abundances, and in the case of Rup\,106 also low n-capture (Eu,
Ba)-element abundances (Gratton {\it et al.}\ 2004). Possibly, they
formed somewhere else at a slower rate, allowing relatively more SN
Type Ia enrichment, but apparently very little by massive
stars. However, their ratio of r-process and s-process element
abundances appears normal, suggesting AGB stars did not play a more
significant role compared to massive stars than what is normally seen.

\subsubsection{Carbon and nitrogen enhancement}

It is not uncommon for cool giant stars in Galactic GCs to exhibit
strong CN bands. Sometimes, a clear bimodality between CN-strong and
CN-weak stars is observed (e.g., van Loon {\it et al.}\ 2007; Kayser
{\it et al.}\ 2008), but all possible combinations occur. In the
field, only weak CN bands occur (but CH can be strong). CN and CH are
often anticorrelated, suggesting that it is N which varies, not
C. The abundance peculiarities are also detected in MS stars, and the
pattern persists further along the evolutionary path, so these stars
must be well mixed, i.e., they have formed from enriched material not
polluted on their surfaces. The bimodality suggests that some stars
may have been affected by N enrichment while others were not, and the
lack of this signature in field stars suggests that the enrichment is
a particular feature of (Galactic) GCs.

Na correlates with (C)N, suggesting a p-capture process perhaps
related to AGB stars of relatively low masses. In 47\,Tuc and
NGC\,6752, neither Mg, Si, Ca, Fe nor Ba appeared to vary (Cottrell \&
Da Costa 1981), but Al also varies at medium--low metallicity and Mg
also at low metallicity, so the N producers were probably slightly
different, depending on their metallicity or mass, if the mass range
of the polluters correlates with the GC metallicity.

The sum of C+N+O generally remains constant. Decressin {\it et al.}\
(2009) found that, as C+N+O increases in rotating massive AGB stars,
nonrotating massive AGB stars are more likely to be responsible for N
enhancement in GCs, but they consider this awkward as one may expect
GC stars to rotate faster, not slower, than in the field. There is
also no obvious reason for a bimodality in the rotation rates, which
might mean that the polluters were never part of the GC.

\subsubsection{The O--Na anticorrelation}

An anticorrelation between O and Na is seen in all Galactic
GCs. [O/Fe] can vary within an order of magnitude, roughly around
[O/Fe] $\approx0$ dex. For [O/Fe] $<0$ dex, [Na/Fe] $\approx0.4$ dex,
but there is a huge spread down to [Na/Fe] $\approx-0.3$ dex for
[O/Fe] $>0$ dex. The explanation for the O--Na anticorrelation is the
production of Na in the CNO cycle, and in associated cycles at higher
temperatures (involving, progressively, Ne, Mg and Al). These cycles
occur in low-mass stars (H shell) as well as intermediate-mass stars
(HBB), so they can give rise to evolutionary as well as primordial
enrichment: MS-star abundances can differentiate between these
scenarios. Rotating massive stars can also produce Na, Al and N
enhancement (Decressin {\it et al.}\ 2007).

\subsubsection{Helium}

He is hard to study spectroscopically, but it has dramatic photometric
effects: for instance, He-rich HB stars will be bluer. In hot-H
burning, He may be correlated with Na and with O depletion. In that
case, one would expect stars on the blue HB to be Na-rich and O-poor,
and vice versa for stars on the red part of the HB. Gratton {\it et
al.}\ (2004) found that while [Fe/H] is higher for hot ($T>11\,500$ K)
HB stars, [He/H] is in fact lower.  Villanova {\it et al.}\ (2009)
found a homogeneous $Y\approx0.25$ for stars in the `safe' range
$8500<T<11\,500$ K, and Catelan {\it et al.}\ (2009) ruled out He
enhancement to explain the HB morphology in the Galactic GC M\,3.

\subsection{The formation of the Galactic globular clusters}

The Galaxy's baryonic halo only contains $\sim10^9$ M$_\odot$, of
which less than a tenth is contained in globular clusters. This is
only $\sim1$\% of the total baryonic mass in the Galaxy. So, there may
not have been more than $10^{10}$ M$_\odot$ of gas involved in forming
it, roughly the amount of gas currently present in the disc. If we
spread this over a volume of, say, $10\times10\times10$ kpc$^3$, we
get a density of 3 particles cm$^{-3}$, similar to that in the disc
and to the atomic--molecular transition in the diffuse ISM. One could
therefore imagine to form a `halo' by instabilities in the diffuse ISM
creating molecular clouds. However, `giant' molecular clouds like
those in the disc do not form GCs. In any case, a density gradient
would be established quickly, concentrating gas into the inner
portions of the hypothetical volume. It is thus attractive to form
both the halo and bulge field and GC populations in the central
regions of the proto-Galaxy, where `hyper' molecular clouds, an order
of magnitude larger and denser than giant molecular clouds, might have
formed.

Concentrating star formation within a small volume naturally explains
a fast and global enrichment, dominated by massive stars, leading to
the observed halo GC age--metallicity relation (Fraix--Burnet et al.\
2009) starting from [Fe/H] $\sim-2.3$ dex. M\,15 is one of the most
metal-poor GCs and unique among these in also showing large
star-to-star variations in r-process elements (at constant
r/s). Perhaps M\,15 was one of the first GCs to have formed, when
conditions were changing fast.

Stellar feedback, possibly in combination with an active galactic
nucleus outflow, would have driven gas into intergalactic
space. Star-formation efficiencies rarely reach 20\%, but are often
considerably lower. Feedback processes resulting from the
star-formation process itself truncate the collapse of the entire
molecular cloud into a myriad of stars. Hence, more than half of the
system's mass might have been ejected. This would have induced the
gravitational unbinding of the central stellar system, with subsequent
migration of its stars and GCs into the halo, a `popping'
proto-Galaxy, only to be retained within the confines of the
dark-matter potential.

The downsizing of clusters over time is intriguing, and suggests that
even the extremely metal-poor halo stars might have formed in a few
superclusters which quickly disintegrated and dispersed, possibly as a
result of dramatic mass loss due to winds and SNe (especially if the
IMF was top-heavy).  A single `hyper'-cluster would be less consistent
with the large star-to-star variations seen at [Fe/H] $<-2$ dex
between r- and s-process enrichment.

A second phase of gas accretion, possibly resulting from the
previously ejected gas cooling and falling back, might then have given
rise to the rapid formation of the bulge and bulge GCs in the nucleus
of the fledgling Galaxy.  This was probably also accompanied by quick
relaxation, given the current kinematics, but not as violent as the
initial phase. These two phases might have coincided with, or indeed
caused, two distinct epochs of re-ionization of the Universe. The disc
formed after that, in a milder manner, and in fact is still seen to be
accreting from the halo and satellite galaxies and forming stars
throughout at 4 M$_\odot$ yr$^{-1}$.

\subsection{Extragalactic populations of star clusters}

Other galaxies may offer different environments and histories in which
to study star clusters and their chemical evolution (cf. Harris
2010). The most accessible of these, the Small and Large Magellanic
Clouds (SMC and LMC, respectively), have formed stars and clusters for
much of their lifetimes and continue to do so. The metallicities of
the young and intermediate-age clusters are lower than of those in the
Galactic disc by factors of $\approx2$--3 (LMC) and $\approx3$--10
(SMC). In the SMC, populous clusters are found at ages of
$\approx6$--7 Gyr (and younger). In the LMC, similar clusters are
found, but not in the 4--10 Gyr interval. This has been linked to
either a lull in cluster formation or more effective cluster dispersal
in the LMC compared to the SMC.

The oldest SMC cluster (and its only GC), NGC\,121, already has a
metallicity [Fe/H] $\approx-1.2$ dex, but then chemical enrichment was
slow and only ramped up by a factor of $\sim3$ in the last few Gyr
(Parisi {\it et al.}\ 2009). Others found a more gradual enrichment,
which seems more plausible given the properties of the populous
clusters of $\sim6$ Gyr old. Piatti {\it et al.}\ (2007) argued for a
different chemical evolution of clusters and the field in the 4--10
Gyr period, but how this would work is not clear. The Da Costa \&
Hatzidimitriou (1998) and Parisi {\it et al.}\ (2009) data possibly
suggest a dip in metallicity around 4 Gyr ago, which is also seen in
the field (Harris \& Zaritsky 2004).  Could this be due to the
accretion of metal-poor gas from the circumgalactic environment, and
could this have triggered enhanced cluster formation in the LMC since
then? Hints at a similar formation of clusters in the Magellanic
Clouds from metal-deficient gas in the past $10^8$ yr was discussed in
van Loon {\it et al.}\ (2005). No\"el {\it et al.}\ (2009) found a
monotonic age--metallicity relation, although they confirm earlier
findings of recent (past few $10^8$ yr) star-formation activity
possibly related to an LMC--SMC encounter. They do not find an old
halo surrounding the SMC. Although consistent with the existence of
just one GC, it begs the question as to what caused the metallicity to
start at [Fe/H] $\sim-1.5$ dex?

The intermediate-age cluster NGC\,419 in the SMC is at a distance of
50 kpc, i.e., 10 kpc in front of the main stellar body of the SMC
(Glatt {\it et al.}\ 2008), and it has a velocity of 188 km s$^{-1}$
(Dubath {\it et al.}\ 1997). This is more than for any SMC cluster in
Parisi {\it et al.}\ (2009) and closer to that of the LMC
($\sim200$--300 km s$^{-1}$). The similarity in age, metallicity and
the properties of its evolved giant stars to the LMC cluster NGC\,1978
was noted in van Loon {\it et al.}\ (2008). Only intermediate-age
clusters are seen around that distance in front of the SMC. The older
clusters concentrate around 60 kpc (Glatt {\it et al.}\ 2008). A link
between these intermediate-age clusters and an origin (of the clusters
or the gas they formed from) in the LMC is tempting.

The Andromeda spiral galaxy (M\,31) also harbours populous
intermediate-age clusters (Barmby {\it et al.}\ 2009), besides a
bimodal GC system that differs from that of the Milky Way in detail
(Galleti {\it et al.} 2009). Several dwarf galaxies in the Local Group
have one or a few (old) GCs associated with them (cf.\ Strader {\it et
al.}\ 2003). Further afield, the GC systems of large elliptical
galaxies are generally found to have bimodal distributions of colours
and metallicity indices, possibly analogously to the Galactic halo and
bulge populations. The bimodality has been interpreted as due to a
past major merger, but there is little direct evidence supporting this
scenario. One might ask why only one major merger occurred, and not
two, or three. Perhaps more systematic investigations of the
properties of GC systems as a function of host-galaxy type will offer
answers.

\section{The end of a paradigm: clusters as composite systems}

Discoveries of dispersion in the colours of giant branches and
star-to-star variations in abundances of N, Ca, etc., already
indicated that stars in Galactic GCs do not have identical
composition. That this might be caused by differences in age and
primordial enrichment has been validated in recent years by accurate
photometry and spectroscopic confirmation in a growing number of GCs,
revealing not only dispersion in stellar properties but also separate,
discrete populations. Although so far found in massive GCs, this is no
longer the exclusive property of the peculiar, most massive GC,
$\omega$\,Centauri (cf.\ McDonald {\it et al.}\ 2009). A summary is
presented in Table 1, roughly in order of decreasing confidence and
cluster mass, $M$.

\begin{table}
\caption{Galactic clusters (possibly) containing multiple populations.}
\begin{tabular}{lccll}
\hline
Cluster(s)    & bulk [Fe/H] & $M$ ($10^6$ M$_\odot$) & Populations & Remarks \\
\hline
\multicolumn{5}{l}{\it Splitting of the main sequence} \\
$\omega$\,Cen & $-1.6$ & 3 & $\geq3$ over 3 Gyr &
              Fe spread \\
NGC\,2808     & $-1.2$ & 1 & 3 within 1 Gyr &
              \\
\multicolumn{5}{l}{\it Splitting of the (sub-)giant branches} \\
M\,54         & $-1.6$ & 2 & 2 &
              Sgr\,dSph nucleus? \\
NGC\,6441     & $-0.5$ & 2 & 2 &
              $\Delta Y\sim0.1$ ? \\
NGC\,6388     & $-0.6$ & 1 & 2 &
              \\
M\,22         & $-1.6$ & 0.6 & 2 &
              Fe spread \\
NGC\,1851     & $-1.2$ & 0.5 & 2 &
              $\Delta Y\sim0$ ? \\
M\,4          & $-1.2$ & 0.2 & 2 &
              bimodal O and Na \\
\multicolumn{5}{l}{\it O--Na anticorrelation} \\
all           & $-2.3$ to $\sim0$ & $\sim0.01$ to 3 & spread (in age?) &
              increases with $M$ \\
\multicolumn{5}{l}{\it N abundances} \\
all?          & $-2.3$ to $\sim0$ & $\sim0.01$ to 3 & often bimodal &
              \\
\multicolumn{5}{l}{\it Horizontal-branch morphology (`second parameter')} \\
all?          & $<-1$ ? & $\sim0.01$ to 3  & unclear &
              not primordial? \\
\multicolumn{5}{l}{\it White-dwarf luminosity function} \\
NGC\,6791     & $+0.4$ & $>0.004$ & 2 ($t\approx8$ Gyr) &
              spread in Na? (He?) \\
\hline
\end{tabular}
\end{table}

The evidence for multiple populations roughly falls into three main
categories: (1) multiple subgiant branches generally indicate discrete
formation epochs separated by a measurable delay, (2) multiple MSs
indicate most likely differences in He abundance, possibly without the
large delays seen in the subgiant-branch splitting, and (3) a spread
in chemical properties of the stars may be due to an extended period
of star formation and continuous chemical enrichment. The distribution
over chemical properties may also show discrete populations. The
presence or absence of subgiant-branch splitting can then constrain
the delay in enrichment between these populations.

\subsection{$\omega$\,Centauri: a freak?}

The photometric detection of discrete populations in the
colour--magnitude diagram (CMD) of the most massive Galactic GC,
$\omega$\,Cen (Lee {\it et al.}\ 1999) added complexity to the
hypothesized star-formation history giving rise to the spread in
metallicity (from [Fe/H] $\approx-1.7$ to $\approx-1.2$ dex), O--Na
anticorrelation, and metallicity-correlated s-process enhancement
observed by Norris \& Da Costa (1995) and Smith {\it et al.}\
(2000). These authors had suggested an extended period of star
formation ($\approx2$ Gyr), with the metal-richer stars forming from
material enriched by massive stars ($>8$ M$_\odot$; $\alpha$ and
Fe-peak elements) and eventually also by contributions from
intermediate-mass AGB stars ($\approx1.5$--4 M$_\odot$; heavy
s-process elements). After having identified a sparse ($\approx5$\% of
all stars), particularly red and dim branch off the main RGB, Pancino
{\it et al.}\ (2000, 2002; see also Origlia {\it et al.}\ 2003) proved
that this `anomalous' RGB-a is the metal-richest population (up to
[Fe/H] $\approx-0.5$ dex) and has lower [$\alpha$/Fe] and higher
[Cu/Fe] ratios indicative of SN Ia enrichment. Stanford {\it et al.}\
(2007) obtained spectroscopic abundance measures in MS stars, which
are unaffected by internal processes. They confirmed a 2--4 Gyr
formation period with a monotonic age--metallicity relation, and that
the C, N and Sr enhancements are primordial (with N enhanced in
metal-richer stars; cf.\ Kayser {\it et al.}\ 2006). Thus, a simple,
sequential chemical-enrichment scenario emerges.

Meanwhile, complications had arisen. A double MS had been discovered
(Anderson 1997; Bedin {\it et al.}\ 2004). Piotto {\it et al.}\ (2005)
then provided spectroscopic evidence showing that the blue MS
corresponds to the intermediate-metallicity population ([Fe/H]
$\approx-1.2$ dex), but the red MS corresponds to the dominant
metal-poor population ([Fe/H] $\approx-1.7$ dex). This could only be
explained if the intermediate-metallicity MS stars were He-enriched by
$\Delta Y\approx0.14$! (Curiously, the metal-richest population may be
He-normal.) Piotto {\it et al.}\ suggested that most of the SN ejecta
from the metal-poor population must have left the system, leaving less
massive stars to create the chemical imprint upon the
intermediate-metallicity population.

Recently, Johnson {\it et al.}\ (2009) distinguished {\em four}
populations, the intermediate-metallicity population splitting up into
slightly metal-richer ([Fe/H] $\approx-1.05$ dex) and more
metal-deficient ([Fe/H] $\approx-1.45$ dex) subpopulations. They found
that, of the metal-poor stars ([Fe/H] $<-1.2$ dex), about half formed
from Type II SN ejecta (similar to disc and halo field stars) but the
other half are also enriched in ejecta from $\approx4$--8 M$_\odot$
AGB stars. The majority of metal-rich stars ([Fe/H] $>-1.2$ dex)
appear to have formed largely from AGB ejecta. They further suggested
that large values of [La/Eu] $>+1$ dex (accompanied by Ba
enhancements) indicate that $\approx25$\% of stars are affected by
binary mass transfer (none of the most metal-poor stars, and more at
higher metallicities). Although these observations can only be
explained with a prolonged formation history, the mildness of Type Ia
SN enrichment favours shorter ($\approx1$ Gyr) rather than longer
($\approx4$ Gyr) timescales.

These four groups relate to those identified in a comprehensive
compilation of data on the MS, subgiant branch and RGB by Villanova
{\it et al.}\ (2007). They traced the [Fe/H] $\approx-1.1$ dex
population into the red MS. However, the brightest (faintest) subgiant
branch belongs to the metal-poor (metal-rich) population. If,
following Villanova {\it et al.}, these are interpreted as measures of
age, then conventional models would assign younger ages to the
metal-poor stars and older ages to the metal-rich stars. Villanova
{\it et al.}\ remained uncertain about the metal-richest ([Fe/H]
$\approx-0.6$ dex) stars on the RGB-a, and there appears to be an
additional (fifth) intermediate subgiant branch which is difficult to
link to just one of the components discussed. They arrive at a rather
curious picture for the make-up of $\omega$\,Cen:
\begin{itemize}
\item {The oldest stars (12--13 Gyr) belong to a trace population
($\approx10$\%)of metal-rich ([Fe/H] $\approx-1.1$ dex) stars;}
\item {Stars of similarly old age, possibly slightly younger (12 Gyr),
belong to an $\approx 20$\% population of metal-poor ([Fe/H]
$\approx-1.7$ dex) stars;}
\item {A much younger population (9--10 Gyr), $\approx30$\% of the
total MS population, consists of considerably He-enriched,
intermediate-metallicity ([Fe/H] $\approx-1.4$ dex) stars;}
\item {The youngest stars (9 Gyr) belong to the dominant
($\approx40$\%), metal-poor ([Fe/H] $\approx-1.7$ dex) population.}
\end{itemize}
Or can these inferred age differences be brought back to within $\ll
1$ Gyr by allowing for variations in the sum of C+N+O and/or He
content?

\subsection{Other clusters: a trend?}

Also a massive GC, NGC\,2808 has revealed three of its MSs, which
probably formed well within a Gyr (Piotto {\it et al.}\ 2007). This GC
is interesting as it has been associated with a stellar overdensity in
the halo, possibly the debris of a disrupted dwarf galaxy. It is also
one of the few coincidences with an H{\sc i} cloud, of $\approx200$
M$_\odot$ (Faulkner {\it et al.}\ 1991), although the association is
uncertain.

Multiple (two) subgiant branches have been found in NGC\,1851 (Milone
{\it et al.}\ 2008), another massive GC and one with both an extended
blue HB and a red clump. It is tempting to relate these two groups of
core-He-burning giants to each of the two subgiant branches and hence
to populations of different age. No split of the MS has been revealed
to date, suggesting that no He enrichment took place between these two
star-formation epochs. Ventura {\it et al.}\ (2009) and D'Antona {\it
et al.}\ (2009) argued, however, that the age difference is not real,
but mimicked by an enhancement in the sum of C+N+O. They held AGB
stars responsible for this, as a correlation is observed with Na, Al,
Zr and La (Yong {\it et al.}\ 2009). It would still be simplest to
explain an enriched population as one that formed subsequently, i.e.,
after some delay, but this delay may be small ($\ll$ Gyr).

Splitting of the subgiant branch has been observed in M\,54, M\,4, the
metal-rich GCs NGC\,6388 and NGC\,6441, and also in M\,22 (Marino {\it
et al.}\ 2009), known to exhibit a spread in [Fe/H] (Da Costa {\it et
al.} 2009). Caloi \& D'Antona (2007) explained the HB morphology of
NGC\,6441 by invoking He enrichment by $\approx0.1$ dex, but no
splitting of the MS has been observed. In M\,3, on the other hand, the
distributions of RR\,Lyrae (HB pulsators) and other HB-morphology
aspects suggest something intricate, but the He spread is constrained
to $<0.01$ dex (Catelan {\it et al.}\ 2009), and no evidence of
multiple populations has been found in CMDs of M\,3. The question of
He enrichment, like that of C+N+O enrichment, is therefore still
open. The two subgiant branches discovered in M\,4 by Marino {\it et
al.}\ (2008) may be related to the bimodal O and Na distribution
observed in that cluster, with the O-depleted stars also being CN-rich
(suggesting enrichment by AGB stars experiencing HBB).

One can be bold, and proceed to explain {\em all} oddities observed in
GCs as due to multiple populations (in the sense of them being formed
in distinct events, either in time, place or both). Is the `second
parameter' (\S 3b) related not just to a difference in age between
GCs, but also to the possible presence of a second population within a
given GC, explaining GCs that exhibit both a red and blue HB? Is the
O--Na anticorrelation caused by stars that occupy different places on
that sequence having formed at slightly different times and thus
sampling slightly different chemical enrichment (or dilution)? The
supersolar metal-rich cluster NGC\,6791---`almost' a GC ---appears to
harbour two populations of WDs (Kalirai {\it et al.}\ 2007): could
this also be related to the cluster having formed stars twice?
NGC\,6791 seems chemically homogeneous, except possibly for a spread
in Na (Carretta {\it et al.}\ 2007{\it b}). Could this be related to
differences in He content, perhaps explaining its He WDs?  Then again,
some GCs lack evidence for a composite nature, e.g., NGC\,6397.

Galactic OCs show no O--Na anticorrelation. They all have roughly
solar [O/Fe] and possibly somewhat higher [Na/Fe] than the primordial
component of the O--Na sequence. [O/Fe] in these clusters decreases
with increasing [Fe/H] as in the disc field stars (De Silva {\it et
al.}\ 2009), simply due to the build-up of Fe from Type Ia SNe. Young
clusters do not display any chemical inhomogeneities (D'Orazi \&
Randich 2009), so there is a fundamental difference between the
formation of OCs and that of GCs, which display a composite nature.

The Magellanic Clouds offer an interesting testing ground for
examining the possible origins of the composite Galactic GCs. No
composite GCs have been uncovered, but it was claimed that some
intermediate-age ($\sim2$ Gyr) populous clusters have additional
populations up to 300 Myr younger: NGC\,1846, NGC\,1806 and NGC\,1783
in the LMC (Mackey {\it et al.}\ 2008), and perhaps more similar
clusters (Milone {\it et al.}\ 2009{\it a}). This has now been shown
to likely result from a spread in stellar rotation rates (Bastian \&
de Mink 2009). Although the CMDs of the affected clusters look
suspiciously similar to those of Galactic composite GCs, the effects
of rotation only show up in the small mass range that happens to be
comprised by the populous intermediate-age Magellanic clusters, also
naturally explaining the lack of `multiple populations' in both
younger and older Magellanic Cloud clusters.

\subsection{The origin of the mixed composition of Galactic globular clusters}

The abundance spreads and patterns provide ample evidence for a
continuous range in stellar content of Galactic GCs, most easily
interpreted as due to an extended period of formation. For several
progressively metal-richer populations in $\omega$\,Cen, the
successive contribution by less massive AGB stars is seen, for
instance, in the growing enhancement of Na at similar relative Fe-peak
abundances and $\alpha$/Fe ratios originating from the early
contribution from Type II SNe (Johnson {\it et al.}\ 2009). The
ubiquitous Na--O anticorrelation must also have been established
during cluster formation (Carretta {\it et al.}\ 2007{\it a}). Field
stars, which do not span an O--Na sequence, show supersolar [O/Fe] at
slightly subsolar [Na/Fe], identical to the corresponding locus in the
O--Na sequence of Galactic GCs (Carretta {\it et al}.\ 2009{\it a};
cf.\ Gratton {\it et al.}\ 2001). Na is followed by N, and both the
primordial and extreme ends of the O--Na sequence are more populous in
more massive GCs (Carretta {\it et al.}\ 2009{\it a,b}). In massive
GCs, stars may have formed over a longer period of time and
consequently also from gas enriched by lower-mass ($<10$ M$_\odot$)
stars.

Enrichment of molecular clouds by massive stars on a timescale of
$\sim10$ Myr has been invoked to explain the difference in metallicity
between the youngest and $\sim10$ Myr-older stars in the Orion
star-forming region (Cunha \& Lambert 1994). But D'Orazi \& Randich
(2009) find 30--55 Myr-old OCs to be chemically homogeneous to within
a few percent. Galactic OCs do not show CN bimodality, an O--Na
sequence or multiple populations. In the 30\,Doradus mini-starburst in
the LMC, multiple populations are seen, including a few Myr-old, few
$\times 10^4$ M$_\odot$ cluster, R\,136. But there is no evidence that
the cluster will develop into one with multiple populations of
different age and metallicity. Neither is there any sign of populous
intermediate-age clusters in the Magellanic Clouds gaining an
additional population of younger, metal-richer stars. The fact that
multiple populations are only seen in old, possibly preferentially
massive, clusters may mean it is intimately related to the unique
conditions under which stars formed in the early history of the
Galaxy, probably in a dense medium near the centre of the
proto-Galaxy.

The chemical differences among discrete multiple populations within
individual Galactic GCs are several tenths of a dex, suggesting bursts
of star formation or other individual events. Some of these bursts are
confirmed, by direct age estimates, to be more than a Gyr apart. The
age resolution in $>10$ Gyr-old GCs is insufficient to reveal
populations that differ by less than several $\times 10^8$ yr, which
is an order of magnitude longer than the lifetime of a molecular
cloud, and sufficient for stars with birth masses of $>3$ M$_\odot$ to
have evolved and chemically enriched the local ISM. There are probably
easier ways of producing subpopulations that only differ by a few Myr
in age and little or nothing in metallicity, than forming a cluster,
letting it mature for one or two Gyr, before, somehow, supplying it
with newly-formed stars. For such extensive age differences, one needs
to invoke the GC having remained part of a larger system which was
gas-rich for at least a few Gyr since the GC formed, or for two
independently formed GCs to have merged.

\subsubsection{Composite clusters as a result of mergers}

The velocity dispersion of the GC system is large, and dynamical
friction is not given much chance to convert velocity differences into
internal kinematic heat: GCs would simply pass through each other with
rather minor consequences.  In the formation stages, when peculiar
velocities are $\sim 1$ km s$^{-1}$, mergers of two GCs are more
plausible. But then, one would expect the GCs to be of very similar
age and composition. Pancino {\it et al.}\ (2007) found a similar
rotation rate for all populations in $\omega$\,Cen, and no radial
differences were found in NGC\,1851 (Milone {\it et al.}\ 2009{\it
b}). Memory may have been erased through dynamical evolution of the
merged system, or mergers take place preferentially involving systems
that share similar kinematics (possibly increasing the effectiveness
of dynamical friction, for instance, in co-rotating systems as opposed
to counterrotating systems), or there never was a merger.

Mergers do not explain an extended period of enrichment either, only
possibly a secondary population, while a third population, i.e., a
third merger, would be highly unlikely. If one could separate the
primary and secondary populations, they would not each look like GCs
in their own right, certainly not the secondary population (there are
no solely N-rich clusters).

\subsubsection{Self-enrichment}

Self-enrichment is an attractive scenario to explain abundance
patterns which are only seen in GCs and not in the field, but how
viable is it? If not through their fast winds, the SN deaths of
massive stars surely remove ISM from the GC. But the winds of
intermediate-mass stars are much slower, in particular in metal-poor
systems, and within the escape velocity of massive, compact GCs.
Could the AGB ejecta on their own have formed the next generation of
stars?  Remember that perhaps as much as half of the stellar content
of massive Galactic GCs might have formed in subsequent epochs
(cf. $\omega$\,Cen, NGC\,6441, M\,22, NGC\,1851).

One can easily estimate that stars in the mass range 2--8 M$_\odot$
yield about half as much mass in the form of ejecta as there would be
in stars in the approximate mass range 0.3--0.8 M$_\odot$. To form
stars from this at about the maximum efficiency, $\sim20$\%, would
thus generate a second generation of stars of about 10\% in number
compared to the first generation (cf.\ Yi 2009). To form an N-enriched
population similar or greater in number than a co-existing, N-poor(er)
population would push this scenario to its limits: it already
comprises a generous range in AGB stars and as a consequence also a
Gyr age difference.

Accreted gas may add to AGB ejecta within the GC before forming
stars. Indeed, Ventura \& D'Antona (2009) confirm that AGB ejecta must
be diluted by pristine gas to reproduce the O--Na anticorrelation. The
additional gas may come from the massive stars in the same generation
of AGB polluters (in which case it would also be enhanced in Fe-peak,
$\alpha$ and r-process elements), or it may be remnant gas from which
that GC population had formed in the first place. Alternatively, it
may come from elsewhere, in which case the enrichment of the second
generation of stars would bear no relation to that of the first
generation. While this might help explain the strange temporal and
chemical `order' of the subpopulations in $\omega$\,Cen, it disagrees
with the absence of star-to-star variations in Fe content in most GCs.

A further challenge for the self-enrichment scenario is that clusters
do not remain gas-rich for long, so it is difficult to envisage AGB
ejecta to accumulate within the cluster over a Gyr or so. No gas-rich
clusters are known except very young ones, and even then the gas
usually surrounds, rather than permeates, the cluster. In young
clusters, SNe---if not fast stellar winds---drive out the gas. Old
clusters have little gas, if any, too, despite having had long to
produce it. Interaction with Galactic halo gas is a likely removal
mechanism (van Loon {\it et al.}\ 2009). So, multiple populations in
GCs are likely to have arisen not from gas accumulation but from gas
accretion, probably quite suddenly in the form of a cooling flow (cf.\
Bekki \& Norris 2006). This is corroborated by the central
condensation of the metal-rich subpopulations in $\omega$\,Cen
(Bellini {\it et al.} 2009). This explains the variety of GCs and the
difficulty to present a unified chemical evolution picture for all
GCs.

It has been suggested that nonconservative mass transfer in massive
binaries could provide enough interstellar gas for subsequent star
formation, and that this gas is enriched in He, N, Na and Al and
depleted in C, O and Mg (de Mink {\it et al.} 2009). Although this can
explain some of the chemical-abundance patterns found in GCs, the
majority of massive stars must participate in this scenario for it to
contribute significantly and it does not explain Gyr delays in star
formation.

\subsubsection{Did globular clusters originate in dwarf galaxies or
dark-matter mini-haloes?}

$\omega$\,Cen, with its retrograde orbit, was suggested to be the
nuclear remnant of a dissolved dwarf spheroidal (dSph) galaxy. M\,54
is, in fact, a massive GC at the heart of the Sgr\,dSph, which is
currently being disrupted in spectacular fashion. Attempts to build a
chemical evolution model based upon this scenario invoke a $10^8$
M$_\odot$ system, of which 1\% survives in the form of the present-day
GC (Bekki \& Freeman 2003; Romano {\it et al.} 2009). The attraction
of a more massive stellar system as the birthplace of GCs is that the
deeper gravitational potential could retain more ejecta, but also that
these ejecta could come from a larger population of stars that never
did nor will form part of the GC. Self-enrichment scenarios indeed
require the GC to be embedded in a larger system (Renzini 2008). The
preferential loss of low-mass, older stars in the early dynamical
phases could then also have caused the comparatively large remaining
further-enriched populations (D'Ercole {\it et al.} 2008).

But there are problems with this origin: dSph galaxies show different
abundance patterns (Geisler {\it et al.}\ 2007): the r-process is
relatively more important in dSphs. Several other GCs in addition to
M\,54 are associated with the Sgr\,dSph, also showing chemical
inhomogeneity: where did they get that from, if they are not
nucleated?  If all GCs had formed in dSph galaxies, enough stars would
have been shed to supply the halo and disc with most of their present
stars. This is clearly not the case for the more enriched disc. The
central concentration of the bulge GCs is also inconsistent with minor
mergers and perhaps the result of star formation triggered by a major
merger (Griffen {\it et al.} 2009).

It has also been suggested that the very distant ($\sim100$ kpc) GCs
might have originated elsewhere, for instance, the metal-poor,
spatially very extended GC NGC\,2419 (van den Bergh \& Mackey 2004) in
which Ripepi {\it et al.}\ (2007) find no spread in metallicity and no
spread in the branches, except an extended HB. MGC\,1, a very remote
GC in the M\,31 halo, may be similar (Mackey {\it et al.}\
2009). Pal\,3 looks like an archetypal GC, with abundance patterns
similar to many other GCs, but different from those in dSph galaxies
(Koch {\it et al.}\ 2009). Curiously, the $\sim10^8$ M$_\odot$
required for the formation and self-enrichment of $\omega$\,Cen is
similar to the potential total mass associated with Pal\,4 and its
H{\sc i} cloud (van Loon {\it et al.}\ 2009). Could the putative
larger systems within which GCs formed be largely dark? A `halo' has
been discovered around NGC\,1851 (Olszewski {\it et al.}\ 2009): could
it have a dark-matter halo too? Could dark-matter haloes be the cause
of the large extent of NGC\,2419 and MGC\,1?  Griffen {\it et al.}
(2009) show that the distant GCs may indeed have kept dark haloes. (GC
dark haloes may only become noticeable beyond the standard tidal
radius.)

\section{Epilogue}

In the briefest of summaries, one might say that above a certain mass
(of the molecular cloud from which it formed, or perhaps rather of an
associated dark-matter component) the formation of a cluster takes
enough time for chemical enrichment to take place during the formation
of subsequent generations of its stars. Such clusters foremostly probe
the conditions of their formation. At lower masses, clusters are
formed essentially instantaneously as a single, homogeneous
population. Such clusters are more unique probes of the star-formation
and chemical-enrichment history of their host (taking into account
cluster dynamics and dispersal).

To make progress, more complete abundance measurements (He!) are
needed, preferably in MS stars (not He), also in the less massive and
more distant GCs and especially the metal-poor ones to make a more
direct connection to the formation of the halo. The interpretation of
such measurements remains problematic: more precise age determinations
are needed, distinguishing formation epochs within GCs and between GCs
differing by $\ll 1$ Gyr. Also, yields of individual elements are not
reliable in the quantitative detail needed to reconstruct the rapid
formation of old Galactic GCs.

Soon, clusters in dwarf and spiral galaxies throughout the Local Group
will be subjected to similar chemical analysis as Galactic clusters
have been, both through photometry and spectroscopy. Environments may
become accessible which resemble the conditions in which the Galactic
GCs we see today were formed, such as those of interacting galaxies,
both nearby and high-redshift (hence historic) starbursts and
star-forming galaxies with ISM metal deficiencies typical of Galactic
GCs, for instance I\,Zw\,18.

As we find that we cannot measure directly what happened in the past,
or that there are too few clusters with which to describe their
population, we may need a theoretical framework to constrain the
freedom of our imagination.  (This may take the form of the ever more
realistic models of cluster and galaxy formation.)

\begin{acknowledgements}
I would like to thank Joana Oliveira for helping me improve the
manuscript, and the editor and referee for their kind patience and
tolerance in enforcing the page limitations.
\end{acknowledgements}

\label{lastpage}
\end{document}